# Critical current density and magnetic phase diagram of $BaFe_{1.29}Ru_{0.71}As_2$ single crystals

**Shilpam Sharma, K. Vinod, C. S. Sundar and A. Bharathi*[*]**

Condensed Matter Physics Division, Materials Science Group,
Indira Gandhi Centre for Atomic Research, Kalpakkam, India - 603102
*corresponding author: *bharathi@igcar.gov.in*

**Abstract:** The critical current density has been measured on single crystals of Ru substituted $BaFe_2As_2$ superconductor at several temperatures and in fields up to 16 T. The magnetisation versus field isotherms reveal the occurrence of a clear second magnetisation peak (SMP) also known as fish-tail effect for both H parallel and perpendicular to c-axis of the crystal. The in-field resistance and magnetisation data are used to put forth a vortex phase diagram. The nature of the vortices has been determined from scaling behaviour of the pinning force density extracted from the $J_C$-H isotherms. The scaled $J_C$ versus reduced temperature behaviour seems to fit to a power law that indicates unambiguously that pinning in this system arises due to the spatial variation in the mean free path, viz. δl pinning.

# I. Introduction

With the discovery of a plethora of Fe Pnictide superconductors[1] the similarities and dissimilarities in the properties of the cuprates and iron-pnictides (FePn) have come under scrutiny[2, 3]. A common feature in the two classes of compounds is the close proximity of antiferromagnetism and superconductivity, with the latter arising by doping charge carriers in the electronically active CuO layers of cuprates or FePn layer of pnictides[1, 4, 5]. In the FePn compounds apart from charge doping, isovalent substitution at either Fe[6, 7] or As[8, 9] site and application of external mechanical pressure[10-12] can induce superconductivity. While the FePn superconductors are similar to the cuprates regarding the high $T_C$, layered structure, strong type II nature[13] with critical current being limited by inter-granular dissipation[13], their properties seem more favourable for applications since they display smaller anisotropy in upper critical field[14-16] larger coherence lengths[17], strong pinning energy and low flux creep.

Successful synthesis of wires of substituted Ba122 superconductors [18, 19] have recently been achieved and therefore optimisation of the current carrying capacity ($J_C$) will be useful. The study of $J_C$ of the type II superconductors is accomplished[20] by the study of M(H) loops in the superconducting state which in addition, is used to elicit important information on the nature of vortex matter in a type II superconductor[21]. Two distinct behaviours have been observed in M(H) isotherms of superconductors (a) the occurrence of peak effect (PE) close to vortex melting, as seen in $YNi_2B_2C$[22] and $NbSe_2$[23] and (b) an small increase, in M within the vortex state suggesting an increased $J_C$, termed the second magnetisation peak (SMP), observed in cuprates[24], $YNi_2B_2C$[22] and more recently in Fe based superconductors[25]. The peak effect that occurs close to $H_{C2}$ has been seen in some of the conventional superconductors such as $Nb_3Sn$[26] and $MgB_2$[27] and is believed to arise mainly due to increased pinning from disordered flux lines prior to melting of the flux line lattice. With the help of imaging and Bitter decoration experiments, the origin of SMP is traced to an increased pinning due to a phase transition in the underlying flux line lattice[28]. Both SMP and PE can also occur for the same system and these have been observed in the $M(H)$ isotherms of the cuprates and borocarbides[22]. In the recently studied Fe based superconductors, only the SMP has been seen for example in $(Ba,K)Fe_2As_2$, $Ba(Fe,Co)_2As_2$, $Ba(Fe,Ni)_2As_2$[16] and recently in $FeSe_{1-x}Te_x$[29]. Since the SMP is seen in a variety of Fe based superconductors it is tempting to conclude that the occurrence of SMP is generic to the M-H behaviour in these set of compounds. A notable exception has been the M-H behaviour in the case of isoelectronic substitution of P at As in $BaFe_2As_{2-x}P_x$ samples [30], where SMP is not seen. It would therefore be interesting to see if the SMP can be observed in M-H isotherms of the $BaFe_{2-x}Ru_xAs_2$ sample in which Fe is iso-electronically substituted by Ru[7].

Here we investigate the magnetisation behaviour of superconducting single crystals of $BaFe_{2-x}Ru_xAs_2$ compound for a Ru fraction of $x=0.71$, in fields up to 16 Tesla. We follow the variation of the critical current density in the Ru substituted $BaFe_2As_2$ single crystals using isothermal $M(H)$ measurements. Our findings point to the presence of a pronounced SMP in $M(H)$ and consequently in $J_C(H)$ for both $H//c$ and $H//ab$ directions. To understand the mechanism of vortex pinning in this system, scaling analysis[31] of normalized pinning force $F_p/F_p^{max}$ as a function of reduced field $h=H/H_{irr}$ was performed. The shape of the variation of $J_C$ versus reduced temperature has been used to demonstrate that variations in the scattering length give rise to flux pinning in this class of superconductors.

**II. Experimental Details**

$BaFe_{2-x}Ru_xAs_2$ ($x_{nominal} = 0.75$, $x_{exact}=0.71$) single crystals were prepared using stoichiometric mixtures of FeAs and RuAs powders along with Ba chunks. The FeAs and RuAs powder precursors were prepared under 35 bar pressure of ultra-high purity argon in a stainless steel chamber by heat treating the intimate mixtures of Fe/Ru and As powders in quartz crucibles in the temperature range of 600 °C to 800 °C for 6 hours[32]. The procedure was repeated twice with an intermediate grinding. For the crystal growth, a stoichiometric mixture of the precursor arsenide powders and Ba chunks were assembled in a helium filled glove box into appropriate alumina crucibles, which was subsequently sealed in evacuated quartz ampoules. These were heated at 50 °C/hour, held at 1190 °C for 24 hours, and then slow cooled to 800 °C at a rate of 2 °C/hour, which was then cooled at the rate of 50 °C/hour up to room temperature. A large numbers of small shiny flat-plate like crystals were found in the crucibles. The powder XRD patterns were recorded using STOE diffractometer operating in the Bragg-Brentano geometry on the powdered single crystals. SEM and EDX analysis were employed to obtain the composition, and Laue pattern of the crystals were recorded to check the crystallinity. In-field resistivity of the sample was measured in a home built 12T magnetoresistance cryostat. For the magnetisation measurements, performed in a Cryogenic Inc. (UK) make vibrating sample magnetometer operating at 20.4 Hz, the crystals were shaped as thin bars to have negligible demagnetisation effects. The dimensions of crystals used for present studies were typically 0.5-0.8 mm long with a thickness of 0.01-0.02 mm. The magnetisation isotherms in fields up to 16T were recorded, with field ramped at 0.5T/min, at several temperatures ranging from 2 K up to 15 K. From these *M(H)* isotherms critical current density $J_C$ as a function of field was calculated using Bean's critical state model[20] from which the vortex pinning force $F_p=\mu_0 H J_C$ was also estimated.

### III. Results and Discussion

#### A. Characterization

The powder XRD pattern of the single crystals, show no impurity peaks and the observed reflections could be satisfactorily indexed on the basis of tetragonal cells (space group *I*4/*mmm*). The *a* and *c*- lattice parameters were found to be 4.0196 Å and 12.7944 Å respectively. These parameters used in conjunction with the earlier XRD and phase diagram data of single crystals[33], suggests that the Ru substitution in $BaFe_{2-x}Ru_xAs_2$ single crystals is ~x=0.71. The EDX analysis indicated a Ru fraction of x~0.7 for the crystals which is consistent with the estimate made from lattice parameter values. This probably suggests that nominal and actual compositions are very close in this method of synthesis.

Figure 1a shows the R(T) plots under magnetic field. The zero field R(T) data indicates a $T_C$ onset of ~20 K and the width of the transition was determined to be ~0.5 K (obtained from temperature difference between the occurrence of 90% to 10% values of the onset resistance values). The superconducting transitions were seen to shift parallel to each other without much broadening under the application of field. Zero-field-cooled and field-cooled (ZFC & FC) magnetisation versus temperature, measured at 0.1T, with *H//c* is shown in figure 1b. The transition temperature ($T_C$) arrived at from the bifurcation point of ZFC and the FC curve is at 20 K. The superconducting volume fraction estimates to be more than 90%, when compared with that obtained for the superconducting volume fraction obtained in a Pb sphere of ~1mm diameter. Thus it is evident that Ru substituted single crystal exhibits bulk superconductivity. The sharp superconducting transition determined by R(T) and large diamagnetic signal at low temperatures imply that crystals are of good quality.

### B. Critical Current density

Figure 2a and 2b show magnetic hysteresis *M(H)* isotherms for fields up to 16 T, applied parallel and perpendicular to the *c*-axis, respectively. *M(H)* loops show a second magnetisation peak (SMP) or a clear fish-tail feature for both field directions and no peak was seen close to the $H_{C2}$, implying the absence of the peak effect. We observed similar fish-tail features in several different crystals from the same batch, cut with different aspect ratios, implying that the observed SMP is not an artefact of the sample geometry. It is also noteworthy that the observed fish-tail for *H//c* direction in the present Ru doped single crystal is similar to that observed for Co and K doped BaFe$_2$As$_2$ single crystals[15, 16, 25, 34]. Shown in figure 2b is the M versus H for *H//ab*. It is observed from the figure that the SMP is seen in this field direction too, but is less pronounced as compared to that for *H // c*. These results are similar to that seen in the Ni-doped BaFe$_2$As$_2$ system[35] whereas SMP was observed only for *H//c* in the Co and K doped BaFe$_2$As$_2$ crystals[15, 16, 25, 34]. It is also evident from the figure 2 that the SMP occurs at lower fields for higher temperatures in each field direction. For a particular temperature the fish-tail effect peak shows up at a lower field for H||c direction as compared to that along H||ab direction

Using the Bean model[20], $J_C(H)$ was determined form the *M(H)* isotherms using *$J_C$ = 20Δm[A\*(1-A/3B)]$^{-1}$*, where Δm is the width of magnetic hysteresis in emu/cm$^3$, *A* and *B* are sample dimensions in cm, with *A < B,* and current density is in A/cm$^2$. Plots of critical current density, $J_C(H)$ of the single crystals in *H//c* and *H//ab* directions are shown in figures 3a and 3b respectively. At 4 K and low fields the value of $J_C$ is ~ 10$^5$ A/cm$^2$, which is well above the value of 10$^4$ A/cm$^2$ expected for applications. The low field $J_C$ of the present Ru doped sample is comparable with that of the electron doped (Co and Ni doped) samples[15, 16, 25], but slightly lower as compared to the hole doped (K doped) sample[16, 34]. The higher $J_C$ in the K-doped samples may be due to their higher $T_C$. It is clear from figure 3 that $J_C$ is very large at very low fields which quickly decreases followed by a broad hump at intermediate fields as a consequence of the SMP in the M-H isotherm. For low temperatures and field strengths larger than 9-10T, the $J_C$ remains nearly constant with increasing field. For both *H // ab* and *H // c*, the SMP shifts to lower fields and becomes more pronounced with increasing temperature. It is also evident from Figure 3 that the rate of decrease of $J_C$ at higher fields is lower for *H//ab*, compared with that in the *H//c* configuration. For all the temperatures, value of $J_C(H // c)$ is higher than for $J_C(H // ab)$ up to a crossover field, after which the $J_C(H \parallel ab)$ takes over the $J_C(H \parallel c)$. This indicates better pinning strength for $H\parallel ab$, compared to $H\parallel c$, at higher fields.

### C. Thermomagnetic history dependence

To look for a possible phase transition in the vortex state in $BaFe_{1.29}Ru_{0.71}As_2$ system close to the SMP in figure 3a, the magnetisation versus field was recorded by field reversal from different FC states at 12K. Using similar measurements close to the peak effect in $CeRu_2$ and $NbSe_2$,[23, 36] it was shown that the peak effect arose as a consequence of a first order phase transition[28]. taking cue from these studies, here too we investigate, if the SMP can arise from a first order transition. The measurement protocol involved field cooling the sample up to 12 K from 25 K under different applied fields and then isothermally measuring magnetisation upon reducing the field to zero. It is evident from the figure for field reversal from 6 T to 2.5 T, the curves merge into one curve. For lower fields the M versus H curves merge onto a different envelop curve. The M(H) trace for field cooling upto 2.5T is marked by an arrow in figure 4. All M(H) curves however merge together below $H_{min}$. Thus it is clear that the sample has two different values of thermo-magnetic cycle dependent $J_C$ for a given field, viz., one for field cooling from 6 T to 2.5 T and another for 2.5 T and $H_{min}$. This history dependent magnetisation, seen from figure 4 suggests the presence of a field driven metastability at 12 K in the Ru doped crystal. This points to the occurrence of a phase change taking place around 2.5T and the thermomagnetic hysteresis could arise from the high field phase getting super-cooled to lower fields, on account of the presence of a first order phase transition.

### D. **Vortex pinning mechanism**

To understand the vortex pinning mechanisms it is illustrative to plot the normalized pinning force density versus applied magnetic field[31]. Shown in fig.5 is the normalized pinning force, $f_p = F_p/F_{pmax}$ plotted as a function of reduced field $h=H/H_{irr}$ for different temperatures. The temperature dependent irreversibility field $H_{irr}$ was determined using $J_C=0$ criterion in the Kramer's plot[37] of $\mu_0 J_C^{0.5} H^{0.25}$ vs. $H$, derived from $J_C$ vs $H$ isotherms (cf. fig.3). It is apparent from the figure 5 that for $H // c$, the normalized curves of fp(h,T) collapse on to a single curve. This when fitted with the Dew-Hughes function[31], $h^p(1-h)^q$ results in p~1.95 and q~2.5 for our Ru doped crystals. According to Dew-hughes model[31], the temperature independent Fp scaling and symmetric Fp(h) curves with a peak at $h_{max}$~0.45 indicates a dense vortex pinning nanostructure. This could result from inhomogeneous distribution of Ru ions, which in turn produces a locally varying order parameter[15].

The $h_{max}$ and p and q values obtained from our data shown in figure 5 is similar to behaviour seen in the Co (electron) doped $BaFe_2As_2$ system[15, 38]. A similar analysis for two different doping levels of potassium (hole) in Ba-122 system by two different groups shows a peak around h~0.43[16] attributed to fluctuating orthorhombic structural domains and also at 0.33[34] which the authors believed to be due to arsenic deficiency, leading to pinning from small-size normal-cores.

In type II superconductors the pinning can be either due to spatial variations in transition temperatures termed as $\delta T_C$ pinning or due to spatial variations in the charge carrier mean free path, termed as $\delta l$ pinning[39]. It has been shown that in the case of $\delta T_C$ pinning, normalized $J_C$ follows $J_C(t)/J_C(0) = (1-t^2)^{7/6}(1+t^2)^{5/6}$, whereas for $\delta l$ pinning $J_C(t)/J_C(0) = (1-t^2)^{5/2}(1+t^2)^{-1/2}$, where $t = T/T_C$ is reduced temperature[39, 40]. Figure 6 shows the plot of normalized $J_C(t)$ for H||c direction. The values of $J_C(t)$ are extracted from $J_C(H)$ plots shown in figure 3. $J_C(t)$ plots are normalized with the $J_C(0)$ values obtained after fitting the data to expression corresponding to $\delta l$ pinning. The theoretical plots of $J_C(t)/J_C(0)$ for the two different pinning mechanisms are shown by lines in figure 6. From the plots of $J_C(t)/J_C(0)$ it is clear that δl pinning fits to the experimental data well and is the dominant pinning mechanism. In the sample under investigation, this can occur due to single vortex pinning at random weak pinning centres created by inhomogeneous distribution of Ru ions. Inset of figure 6 shows a similar analysis performed on a different crystal having slightly lower Ru concentration (x~0.6). Here again the data fits better to the δl pinning scenario. It is noteworthy that a dominant δl pinning has also been reported for $FeSe_{0.5}Te_{0.5}$ [29] system.

To understand the kind of vortex phases existing above and below fish-tail effect field, the critical state magnetic relaxation over a period of time for different fields were studied at 12K. A logarithmic relation rate S = -|d ln M/d ln t| describes the flux creep but no single form describes the relaxation phenomenon for all $J_C$ and field[25]. Figure 7 shows the field dependence of relaxation rate *S* measured at 12K above and below the second peak field along with M(H) loop at the same temperature. In all the measurements, the zero field cooled sample was subjected to a high field much beyond the fish tail region. After forming the vortex state at a higher field the relaxation in magnetisation was studied for different fields above and below the fish-tail region. Different time dependencies of M(t) were found above and below the fish-tail peak field. *S(H)* initially decreases with increasing field up to fish-tail peak maxima $H_{SP}$ and thereafter increases with field. This behaviour is similar to what was observed in Co-doped $BaFe_2As_2$ single crystals[25]. Initial fall of *S* with increasing field is in agreement with weak collective pinning creep model but its subsequent increase with field cannot be explained using this model. Thus collective creep model is not applicable above $H_{sp[25]}$. An increase in the relaxation rate with field can be explained by plastic creep model[25, 41]. It was understood that a change from collective to plastic creep is responsible for the fish-tail shape of the isothermal M(H) loops[25].

### E. Vortex Phase diagram

Based on the above studies of magnetisation and its relaxation, a phase diagram of Ru substituted BaFe$_2$As$_2$ system for $H//c$ direction is shown in figure 8. $H_{min}$ and $H_{SP}$ represent the field values at valley after low field peak and a second peak at higher fields in the magnetisation isotherms (cf. fig. 2). Resistive transitions as a function of temperature under different magnetic fields shown in fig 1a were used to calculate upper critical field ($H_{C2}$) and irreversibility field ($H_{irr}$). $H_{C2}$ and $H_{irr}$ were defined by 90% and 10% of normal state resistivity criterion respectively. The curves for $H_{SP}$ and $H_{min}$ fit very well to power law of the form $H_{SP}, H_{min} = H_{SP}, H_{min}(0) (1-T/T_C)^n$, with $H_{SP}(0)$=8.27T and $H_{min}(0)$=1.63T. The value of exponent $n$ is found to be 1.27 and 1.24 for $H_{SP}$ and $H_{min}$ respectively. The exponents seem to agree with those obtained in the FeSe$_{0.5}$Te$_{0.5}$ system[29] but the values of $H_{SP}(0)$, $H_{min}(0)$, and exponents are low as compared to the K-doped Ba-122 system[34]. Another noteworthy feature of the phase diagram shown in figure 8 is that the irreversibility line is very close to the $H_{C2}$ curve. The irreversibility line being very close to $H_{C2}$ is very important feature for a material to be technology worthy as irreversibility line demarcates the field at which vortex flow is unpinned and magnetic irreversibility sets in. In contrast other electron (Ba(Fe,Co)$_2$As$_2$ and Ba(Fe,Ni)$_2$As$_2$) and hole doped ((Ba,K)Fe$_2$As$_2$) compounds of Ba-122 family[16, 25, 34] show significant difference between $H_{irr}$ and $H_{C2}$ lines. A delineation between two different regimes of vortex dynamics[25] marked as phase I (collective creep) and phase II (plastic creep) can be seen separated by the $H_{SP}$ line in the phase diagram.

### IV. Summary and Conclusions

Single crystalline samples of BaFe$_{2-x}$Ru$_x$As$_2$ (x$_{nominal}$=0.75, x$_{exact}$= 0.71) were synthesised using stoichiometric amounts of FeAs, RuAs powders and Ba Chunks. Since no excess flux was used for growth, the crystals were found to be phase pure with little or no impurity inclusion. The powder XRD, Laue, EDX and also the sharpness of superconducting transitions provides evidence of the purity of single crystals. Critical current density of the compound is around ~$10^5$ A/cm$^2$ which is comparable to other electron doped systems of nearly same $T_C$. A second magnetisation peak was observed in both the directions of applied field and can be explained on the lines of Co-doped Ba-122 system to be originating from a crossover from one kind of vortex dynamics to other. Based on the in-field resistivity vs temperature and M(H) measurements, it is suggested that for Ru doped BaFe$_2$As$_2$ system the irreversibility field lies very near the upper critical field. From different behaviours of M(H) curves recorded by reversal of field from a vortex state formed by field cooling the sample to 12K above and below second peak, it is clear that there are different values of $J_C$(H) above and below the second magnetisation peak and thus the critical currents are thermo-magnetic cycle dependent. This suggests that the phase change in underlying vortex lattice could be first order, with associated hysteresis. When cooled from higher fields the $J_C$/magnetisation is higher probably due to additional pinning from kinetically arrested glassy vortex or liquid state or super-cooled state. At 12K and below 2.5T the phase contributing to $J_C$ is the equilibrium vortex phase alone. Field dependent relaxation measurements at 12 K suggest a change of flux pinning from collective creep to plastic creep regime. The temperature-independent scaling behaviour of the normalized pinning force density suggests one dominant pinning mechanism. Similar to Co-doped system[15], the symmetric $F_p$(h) curves with a peak at h~0.45 may imply a dense vortex pinning originating from the local variations in order parameter with evidences of δl pinning as the dominant pinning mechanism in the BaFe$_{1.29}$Ru$_{0.71}$As$_2$ system.

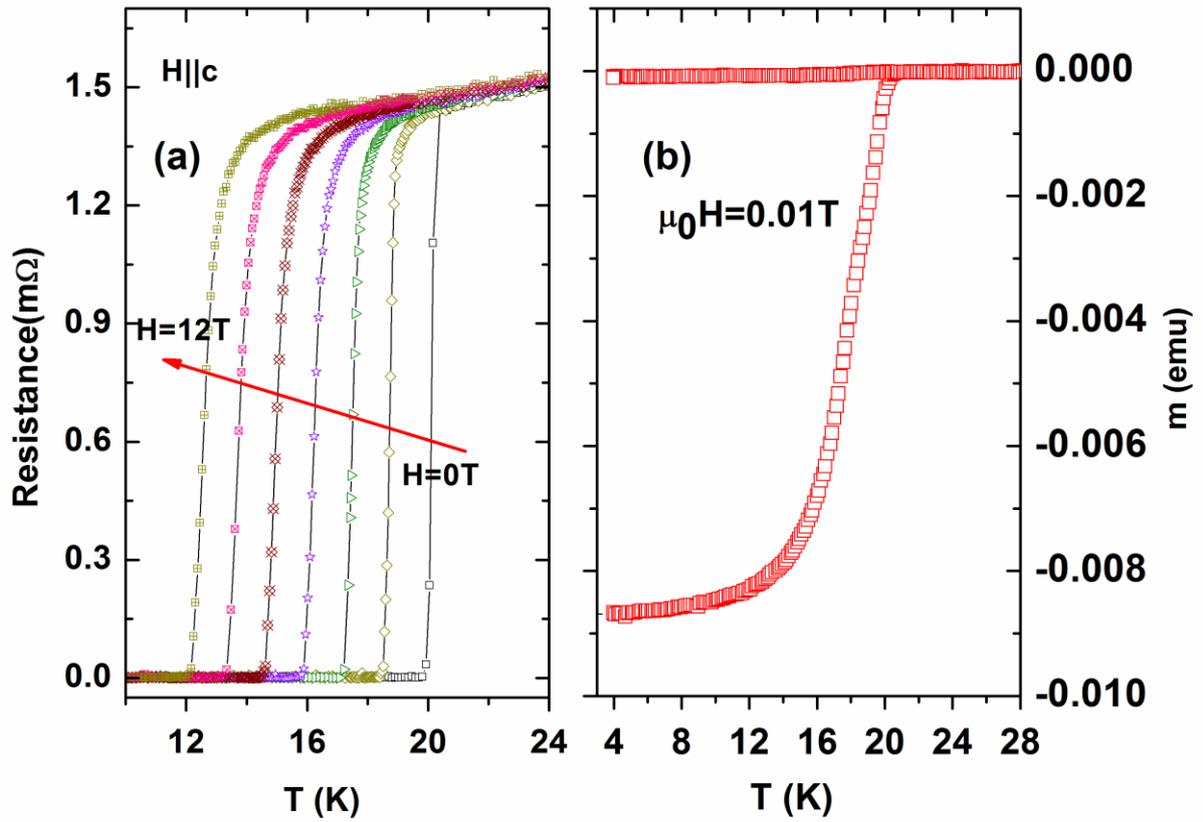

**Figure 1a:** In-field superconducting transitions shifts parallel to each other. **Figure 1b:** M(T) plot at 0.01T shows bulk superconductivity and the sharp superconducting transition points to pure crystals.

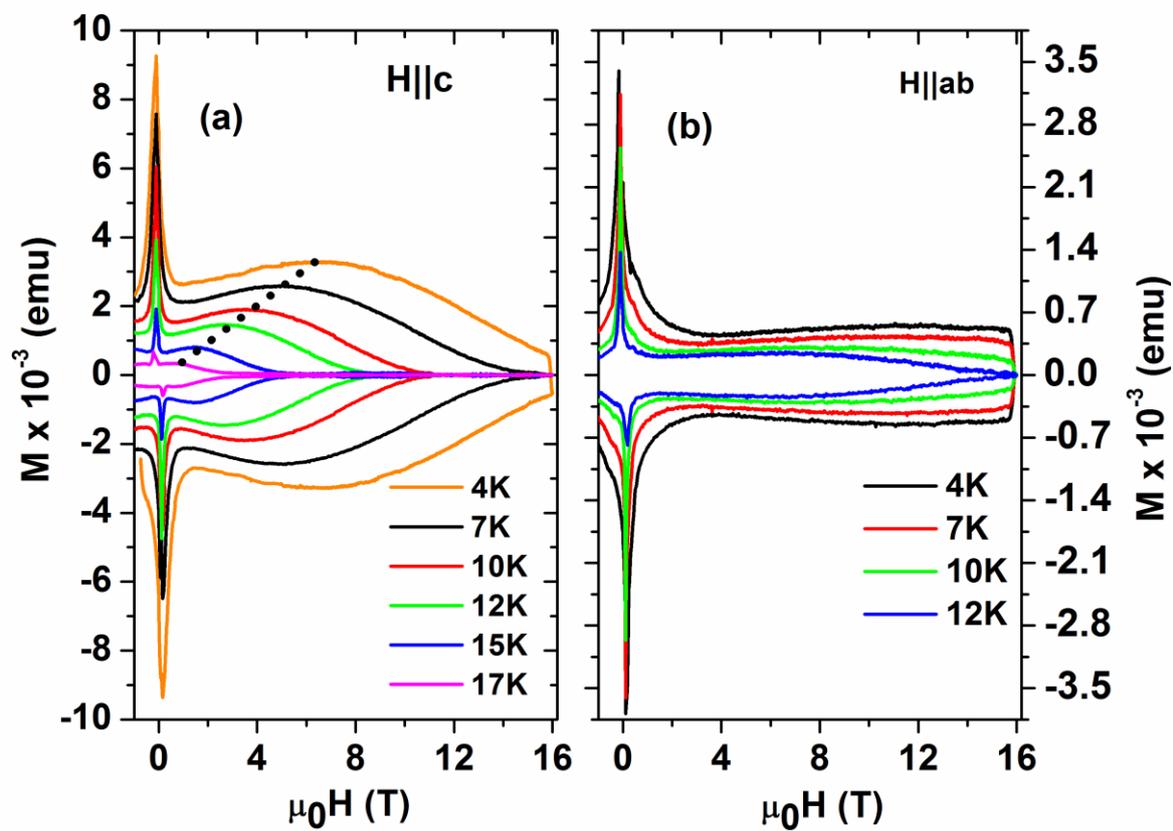

**Figure 2a:** M(H) isotherm for H||c direction shows second-peak effect at a field much lower than $H_{C2}$. **Figure 2b:** M(H) isotherm for H||ab direction also shows a feeble second-peak effect.

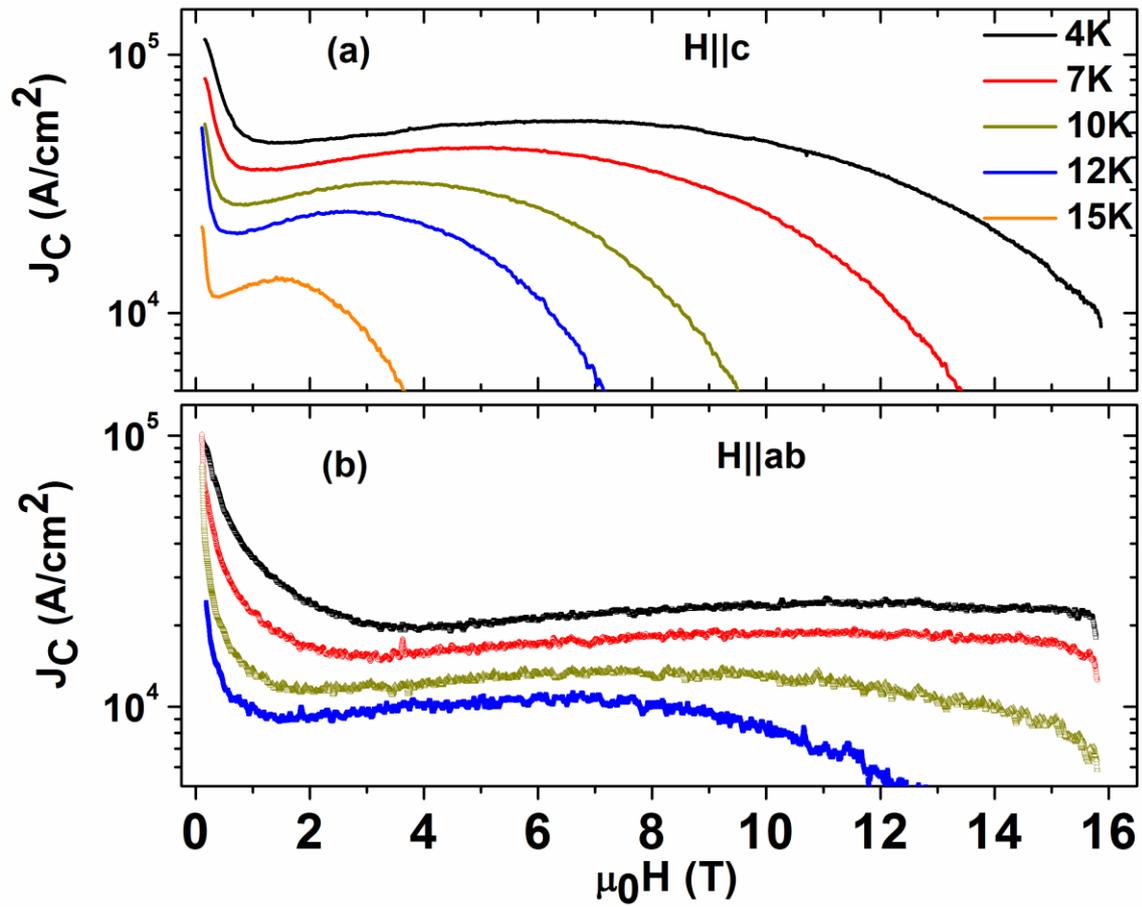

**Figure3a:** Critical current density in H||c direction shows pronounced fish-tail peak. **Figure3 b :** Critical current density in H||ab direction remains weakly dependent on field.

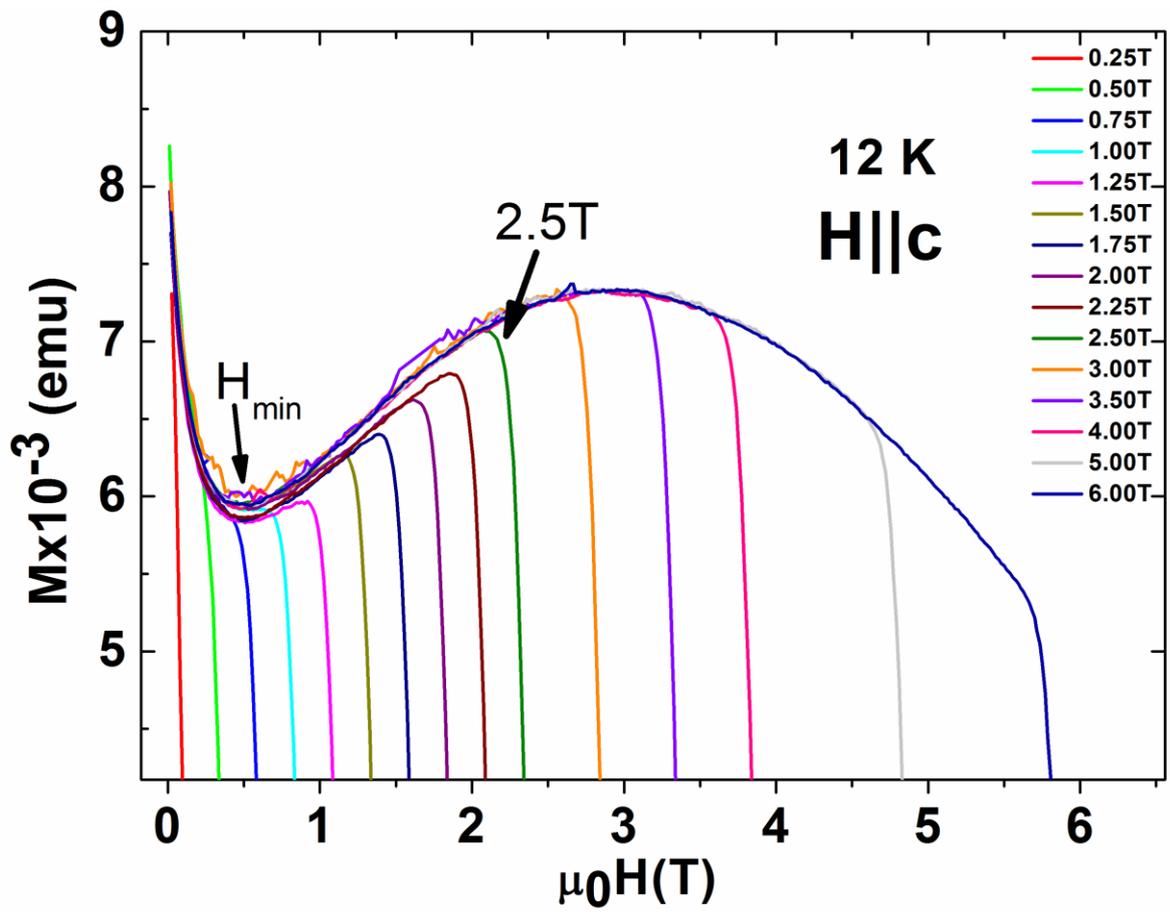

Figure 4 : M(H) curves recorded for field reversal from different vortex states obtained by field cooling from normal state (25 K) upto 12 K in different external magnetic fields indicated. Thermo-magnetic history effects are clearly seen for field cooled states below and above 2.5 T (see text).

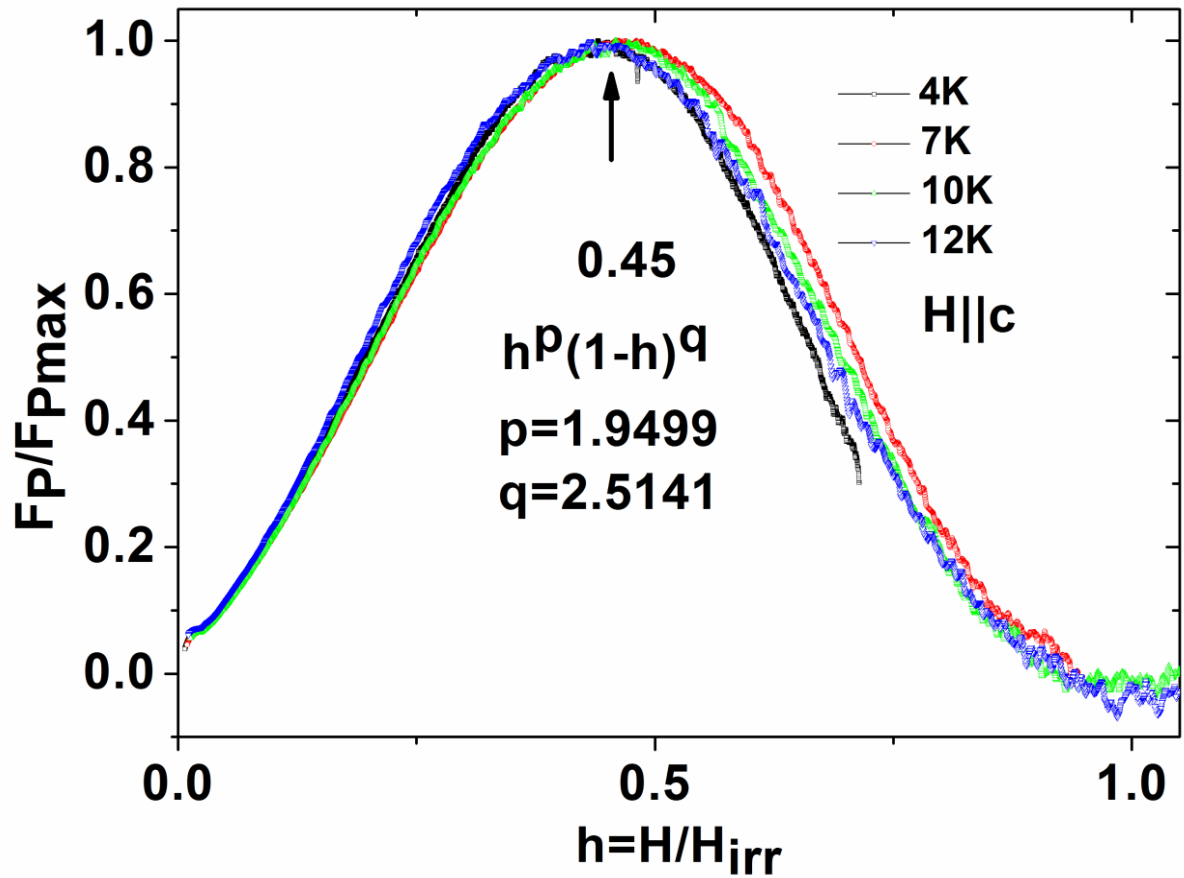

**Figure 5:** Temperature independent scaling analysis of normalized pinning force density. Scaling feature is suggestive of the presence of dense pinning centres due to inhomogeneous Ru ion distribution.

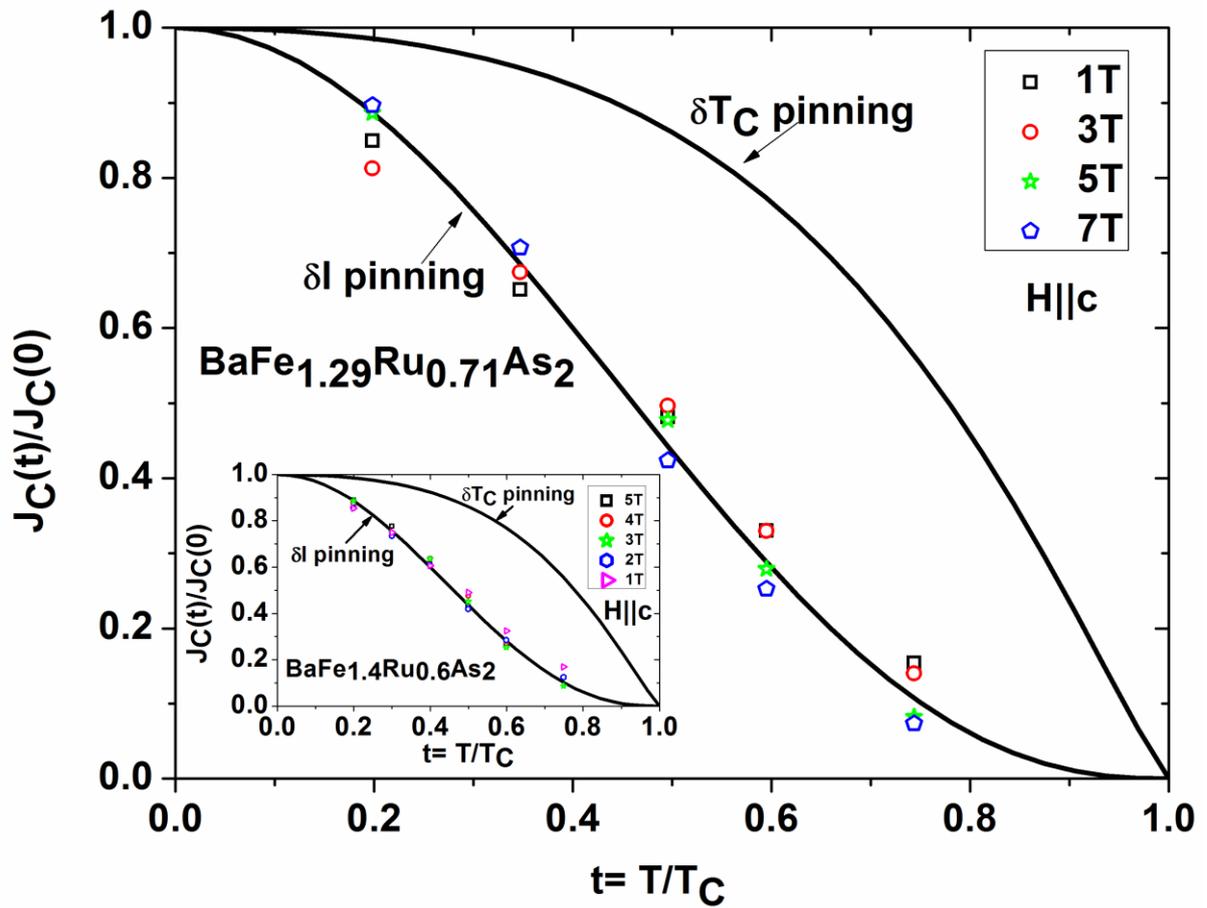

Figure 6: Normalized $J_C(t)$ data in a single crystal of $BaFe_{1.29}Ru_{0.71}As_2$ for H parallel to c axis. The solid lines correspond to the expected variation due to $\delta l$ and $\delta T_C$ pinning (see text for details). Inset: Normalised $J_C(t)$ data for $BaFe_{1.4}Ru_{0.6}As_2$ single crystals.

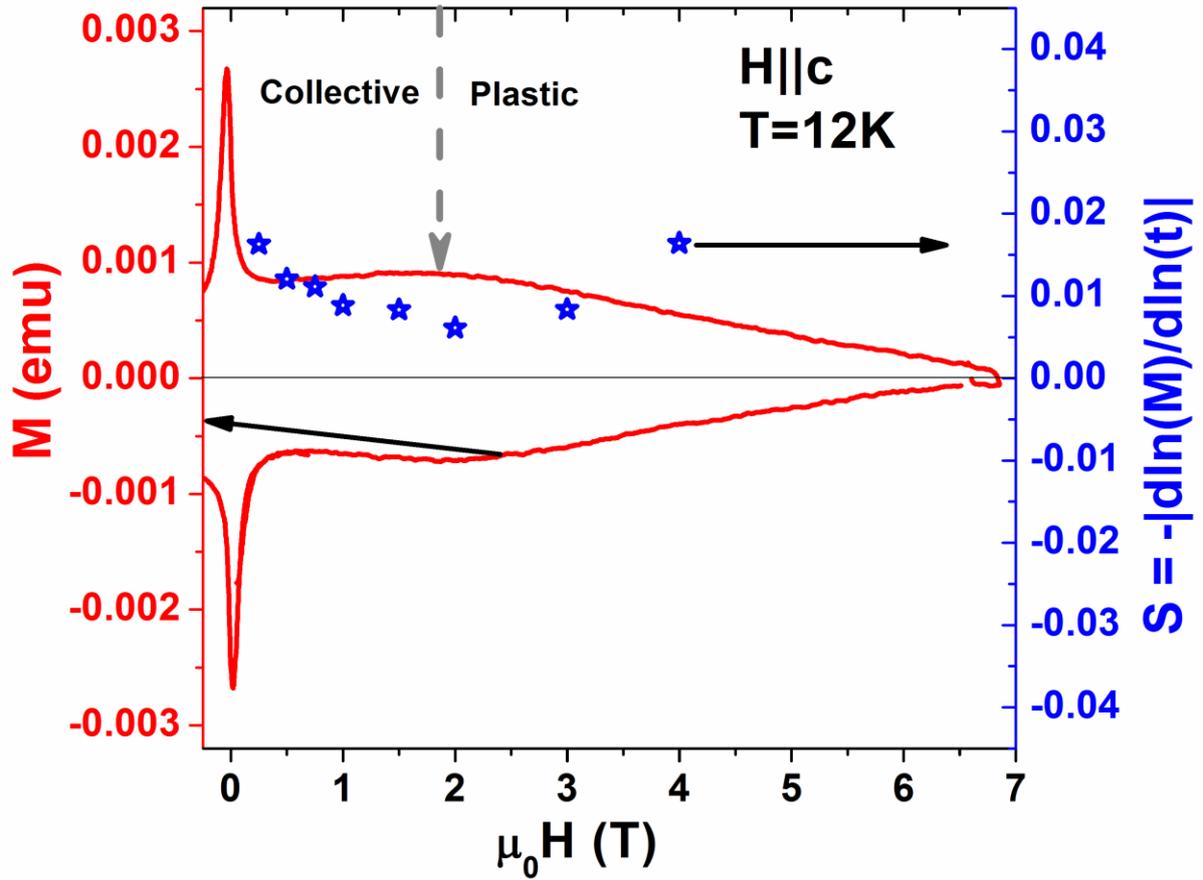

**Figure 7:** Field dependence of magnetisation relaxation rate 'S' (stars) shown along with M(H) loop measured at 12 K for H parallel to c. The arrow indicates the peak that delimits the collective and plastic pinning regimes in the H-T phase diagram.

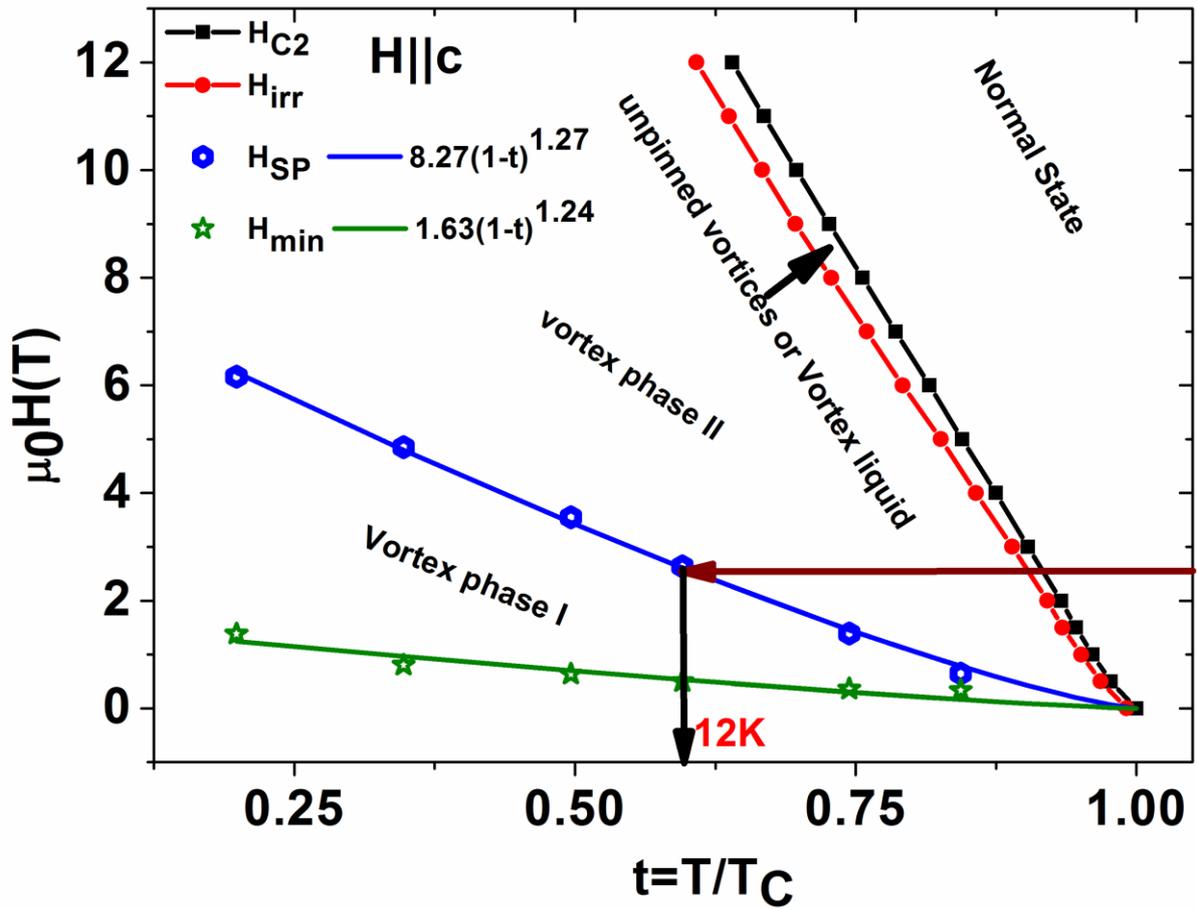

Figure 8: Detailed vortex phase diagram for BaFe$_{2-x}$Ru$_x$As$_2$ superconducting single crystal shows a very small region of unpinned vortices. Arrow marks H=2.5T and T=12K where sample shows a change in thermomagnetic history dependent J$_C$ value.